\documentclass[%
 reprint,
nofootinbib,
 amsmath,amssymb,
 aps,
]{revtex4-1}
\usepackage{booktabs}
\usepackage{graphicx}
\usepackage{dcolumn}
\usepackage{bm}
\usepackage{lipsum}
\usepackage[utf8]{inputenc}
\usepackage{verbatim} 
\usepackage{float}

\begin{document}

\title{Eikonal quasinormal modes, photon sphere and shadow of a charged black hole\\
in the 4D Einstein-Gauss-Bonnet gravity}

\author{Jose Miguel Ladino}
 \email{jmladinom@unal.edu.co}
\affiliation{Universidad Nacional de Colombia. Sede Bogotá. Facultad de Ciencias. Observatorio Astronómico Nacional. Ciudad Universitaria. Bogotá, Colombia.
}

\author{Eduard Larra\~{n}aga}
 \homepage{ealarranaga@unal.edu.co}
\affiliation{%
 Universidad Nacional de Colombia. Sede Bogotá. Facultad de Ciencias. Observatorio Astronómico Nacional. Ciudad Universitaria. Bogotá, Colombia.}%

\date{March, 2023}

\begin{abstract}
In this work, we investigate the relationship between the geometrical properties, the photon sphere, the shadow, and the eikonal quasinormal modes of electrically charged black holes in 4D Einstein-Gauss-Bonnet gravity. Quasinormal modes are complex frequency oscillations that are dependent on the geometry of spacetime and have significant applications in studying black hole properties and testing alternative theories of gravity. Here, we focus on the eikonal limit for high frequency quasinormal modes and their connection to the black holes geometric characteristics. To study the photon sphere, quasinormal modes, and black hole shadow, we employ various techniques such as the WKB method in various orders of approximation, the Poschl-Teller potential method, and Churilova's analytical formulas. Our results indicate that the real part of the eikonal quasinormal mode frequencies of test fields are linked to the unstable circular null geodesic and are correlated with the shadow radius for an Charged Einstein-Gauss-Bonnet 4D black hole. Furthermore, we found that the real part of quasinormal modes, the photon sphere and shadow radius have a lower value for charged black holes in 4D Einstein-Gauss-Bonnet gravity compared to black holes without electric charge and those of static black holes in general relativity. Additionally, we explore various analytical formulas for the photon spheres and shadows, deducing an Churilova's approximate formula for the black hole shadow radius of the Charged Einstein-Gauss-Bonnet 4D black hole, which arises from its connection with the eikonal quasinormal modes.

\end{abstract}


\maketitle

\section{Introduction}

Quasinormal Modes (QNMs) are a distinguishing characteristic of Black Holes (BH) that describe their damped oscillations over the spacetime in response to external perturbations \cite{Kokkotas1999}. The frequencies of these QNMs of BHs are complex numbers, with the real part corresponding to the frequency of the oscillation and the imaginary part corresponding to the rate at which the amplitude of the oscillation decays. QNMs have several important applications in the research of BHs, such as studying their surface gravity and horizon area, stability, the detection of gravitational waves, and their geometrical properties, such as their mass, spin, and electric charge. There are various techniques used to compute and calculate QNMs, including the WKB method, the continued fraction method, the time-domain integration method and much more. In Fact, the knowledge about QNMs could also have potential implications  in other fields such as astrophysics, cosmology, and high-energy physics, like testing General Relativiy (GR) and alternative theories of gravity \cite{Konoplya2011}. In the same way, eikonal QNMs are a specific type of oscillations that occur in the eikonal limit, which is particularly useful for studying the high frequency QNMs and its applications. In this limit, the real part of the QNM frequency is directly related to the unstable circular null geodesics of the BH, which is useful for understanding the connection between the QNMs and the BH geometric properties, e.g. the real part of the QNM frequency is a monotonically increasing function of the spin BH and that the imaginary part of the QNM frequency is a monotonically decreasing function of the spin BH \cite{Konoplya2017}. Another very interesting feature about BHs is its shadow, a darkness region in the vicinity caused by the bending of light by the BHs gravity. The exact shape and size of a BH shadow also depend on the geometrical properties of the BH, as well as the properties of the environment surrounding the BH, such as the distribution of matter and the presence of other objects, e.g. the presence of dark matter affects the QNM frequencies and the shadow radius of the BH, and these two quantities are related in the eikonal limit, so that, the effect of dark matter on QNMs and the shadow radius is stronger for rotating BHs compared to non-rotating BHs \cite{Jusufi2020}.

The BH solution in the Einstein-Gauss-Bonnet (EGB) theory has been derived from various modified gravity theories through different approaches. The initial BH solution in EGB theory was found by Boulware and Deser in 1985 \cite{bolware1985} for dimensions $D \geq 5$, as the GB term does not affect the gravitational dynamics in $D=4$. However, later studies, such as those by Tomozawa in 2011 \cite{tomozawa} and Cognola et al. in 2013 \cite{Cognola2013}, discovered that the GB term could have a non-trivial contribution to spacetime when $D=4$ through regularization and dimensional reduction techniques. According to Lovelock's theorem, EGB gravity is only introduced in $D \geq 5$, as the GB term does not contribute dynamically in lower dimensions \cite{Bousder2021(2)}. This led Glavan and Lin in 2020 \cite{G2020} to propose a rescaling of the coupling constant to obtain a contribution to gravitational dynamics in $D=4$. From this moment, this BH solution in a EGB 4D gravity has been studied intensively and is considerably a subject of ongoing research. For example, this BH solution in EGB theory was  explored subsequently in 2020 \cite{Fernandes2020} by Fernandes, who studied its coupling with both BH electric charge and anti-de Sitter space. Despite the fact that the solution proposed by BH in \cite{G2020} was obtained in a simple and novel way, the model used there was strongly criticized. Several later studies have shown that the method used to find the solution was neither consistent nor well-defined \cite{Gurses2020,GursesNew,Arrechea2020New,Fernandes2022}. However, several subsequent studies have obtained this same solution and clarified that it was not really new. This and other very similar solutions can be deduced from different well-defined approaches and modified gravity theories \cite{Fernandes2022, Cai2010New, Cai2014New, LuNew, HennigarNew, FernandesNew, Aoki2020, Ghosh2021}. Anyway, as noted in \cite{Fernandes2022}, the EGB 4D theory of gravity can be interpreted as an extension of GR with quadratic curvature corrections, yielding interesting implications for BHs, cosmology, and weak-field gravity.

The study of eikonal QNMs in BH solutions of GR and its connection with the photon sphere has been previously explored in \cite{Li2021}. These studies have also been extended to other alternative theories of gravitation, such as Scalar Gauss-Bonnet gravity in \cite{Bryant2021}, Einstein-dilaton-Gauss-Bonnet BHs in \cite{KonoplyaEdGB2019}, dynamical Chern-Simons gravity in \cite{Glampedakis2019}, Rotating Loop Quantum BHs in \cite{Liu2020}, string-corrected D-dimensional BHs in \cite{Moura2021} or for deformed Schwarzschild BHs in \cite{Chen2022}, to give some recent examples. In adittion, The QNMs of BHs in EGB 4D gravity is a topic of ongoing research, some of this investigations have been focusing on the shadows and photon spheres with spherical accretions, on the correlation between the shadow of a BH and its eikonal QNMs, as well as the effect of the Gauss-Bonnet (GB)  coupling constant, $\alpha$, on these properties \cite{Konoplya2020, Liu2022, Churilova2021-2, Zeng2020}. Later, more research has been carried out on extended and more complex versions of this EGB 4D BH type solution. A number of studies have been given lately, including the investigations of eikonal QNMs and greybody factors in asymptotically de Sitter spacetime \cite{Devi2020}, the examination of the shadow of rotating EGB 4D BHs \cite{Kumar2020-5}, the study of null geodesics and shadow of EGB 4D BHs surrounded by quintessence \cite{Heydari2022}, the analysis of entropy, energy emission, QNMs and deflection angle of EGB 4D BHs with nonlinear electrodynamics \cite{Kruglov2021} and the investigation of QNMs of an EGB 4D BH in anti-de Sitter space \cite{Churilova2021}. In the same way, several studies have been conducted on electrically charged BHs in EGB 4D gravity, each focusing on different aspects of their behavior. In \cite{Kumar2020} gravitational lensing is studied, in \cite{Atamurotov2021} particle motion and plasma behavior are examined, in \cite{Zhang2020} superradiance and stability of the solution are discussed and in \cite{Zhang2021} the connection between phase transition and QNMs is explored. Actually, this research trend has been given in mostly general solutions. For example, a recent publication delved into the properties of rotating charged BHs in in EGB 4D gravity, such as the examination of photon motion and its shadow \cite{Papnoi2022}, an analysis was also made of the characteristics of charged EGB 4D BH in anti-de Sitter space, including their shadow, energy emission, deflection angle, and heat engine properties \cite{Panah2020}, as well as the investigations of the instability, QNMs and strong cosmic censorship of charged EGB 4D BH in de Sitter space under charged scalar and electromagnetic perturbations in \cite{Liu2021, Mishra2020}. Notably, by merging the findings from previous investigations, a more comprehensive understanding of these variations in modified theories of gravity can be achieved.

A recent work in \cite{Chen2021} explored the correspondence between the shadow and the QNMs of the scalar field around a charged EGB 4D BH, using the $6^{th}$ order of the WKB method. In this work, we will extend the findings of this study by utilizing additional approaches and methods to explore the relationship between the photon sphere, the BH shadow, and the eikonal QNMs of scalar and electromagnetic field perturbations for a charged EGB 4D BH.

This paper is organized as follows: In Section II, we present the electrically charged BH in EGB 4D gravity, briefly introducing the BH metric background, their horizons, and particular limit cases. In Section III, we show the theory behind the QNMs of scalar and electromagnetic field perturbations, providing the corresponding master wave equations and discussing  some of the principal semi-analytical methods to calculate the QNM frequencies, such as the WKB approximation approach and the Poschl-Teller potential method. Then, in Section IV, we discuss some eikonal QNMs approaches, including a recently proposed analytical formulation that approximates the frequencies at this limit. Later, we analyze and apply these methods on the eikonal QNMs of a charged BH in EGB 4D gravity, looking at the effect of the geometric parameters of the BH on these.  Afterwards, in Section V, we study the photon sphere of a charged BH in EGB 4D gravity and its particular limit cases, illustrating the correspondence between eikonal QNMs and null geodesics, and revealing the effect of the geometric parameters of the BH on this. Thereafter, in Section VI, we investigate the shadow of a charged BH in EGB 4D gravity and its connection with their eikonal QNMs frequencies, sharing an analytic and approximate formula for the shadow and comparing it with all the results given by other methods and again, the effect of the geometric parameters of the BH in their shadow. Finally, in Section VII, we summarize some conclusions.

\section{The Charged Black Hole in the 4D Einstein-Gauss-Bonnet Gravity}

\subsection{The Spacetime Background}
The gravitational theory of EGB in D-dimensional spacetime coupled with an electromagnetic source is described by the action \cite{Chen2021}
\begin{eqnarray}
S&= & \frac{1}{16 \pi G} \int d x^D \sqrt{-g}\left[R-F_{\mu \nu} F^{\mu \nu}\right. \\
&&\left.+\alpha\left(R^2-4 R_{\mu \nu} R^{\mu \nu}+R_{\mu \nu \beta \gamma} R^{\mu \nu \beta \gamma}\right)\right], \notag
\label{eq1}
\end{eqnarray}
where $R$ is the scalar curvature, which corresponds to the well known  Einstein-Hilbert contribution, $R_{\mu \nu}$ and $R_{\mu \nu \beta \gamma}$ are the Ricci and Riemann tensors respectively, $\alpha$ is known as the GB coupling constant and $F_{\mu \nu}$ is the electromagnetic tensor given by
\begin{equation}
F_{\mu \nu}=\partial_{\mu} A_{\nu}-\partial_{\nu} A_{\mu},
\end{equation}
with $A_{\mu}$ being the quadripotential. $\alpha$ has dimensions of $[\text{length}^2]$ and some authors take it between $-8M^2 < \alpha < M^2$ \cite{Konoplya2020}. Nevertheless, it is usual to consider the simple constrain $\alpha>0$ \cite{Ghosh2021}, since it has been shown that for $\alpha<0$ the BH solution might not be valid for small distances \cite{Kumar2020-5}. In this work, we will use the assumption $\alpha>0$.

EGB gravity describes quadratic corrections to the curvature tensors from Lovelock's gravitational theory, but is also obtained in the low-energy limit of string theory, in which $\alpha$ can be interpreted as the inverse stress of the string and is defined only with positive values \cite{Kumar2020-5}. Actually, the BH solution in EGB theory has already been deduced from various modified gravity theories. It was initially obtained in 1985 by Boulware and Deser in \cite{bolware1985} for the cases where $D \geq 5$, since the GB term does not contribute to the gravitational dynamics when $D=4$. Then, Tomozawa in 2011 \cite{tomozawa}, through a regularization procedure, found that there could be, from a quantum perspective, a non-trivial contribution of the GB term to space-time in the case where $D=4$ . Later in 2013, Cognola et al. \cite{Cognola2013}, through another process of regularization to EGB gravity, managed to perform a dimensional reduction under the Lagrangian formulation, finding again the BH type solution for the case in which $D=4$. According to Lovelock's theorem, the gravity of EGB is only introduced in cases where $D>4$ because, for smaller dimensions, the term with the GB coupling would not contribute dynamically \cite{Bousder2021(2)}. In 2020, Glavan and Lin \cite{G2020} used this formalism to obain the same BH solution in EGB 4D gravity by simply proposing a rescaling of $\alpha$, obtaining a contribution to the gravitational dynamics in the case  $D=4$. This rescaling is $\alpha \rightarrow \alpha/(D-4)$.
However, several studies have shown that the proposal to rescale the GB coupling parameter, as proposed in \cite{G2020}, is incorrect. This is easily suspected by the fact that the lagrangian action diverges and is not well-defined in the 4-dimensional limit, disregarding Lovelock's theorem \cite{Gurses2020,GursesNew,Arrechea2020New,Fernandes2022}. Consequently, it has been shown and clarified that the same solution can be obtained from different approaches and theories of modified gravity that are well-defined. Therefore, it is not an entirely novel solution. To obtain coherent versions of the theory of BH solutions of EGB 4D, alternative regularization processes have been applied. These include scalar-tensor theories from conformal regularizations \cite{HennigarNew,FernandesNew}, the Kaluza-Klein regularized reduction \cite{LuNew}, and other formalisms related to theories of gravity such as semi-classical or higher dimensional \cite{Fernandes2022,Cai2010New,Cai2014New,Aoki2020,Ghosh2021}.
This undoubtedly shows that this BH solution has been of great interest in recent years.

The spacetime under consideration is assumed to be static and spherically symmetric, and is described by
\begin{align}
ds^{2}  = & -f\left(r\right)dt^{2} +  f\left(r\right)^{-1} dr^{2} + r^2d\Omega_D^2,
\label{eq:Metric}
\end{align}
where, if $D=4$, we have that $d\Omega_D^2 \equiv  d\theta^{2}+\sin^2 \theta d\phi^{2} $ and
\begin{equation}
A_{\mu }=-\frac{Q}{r}dt,
\end{equation}
so the solution of an electrically charged BH in EGB 4D gravity will take the form
\cite{Chen2021}
\begin{equation}
f^{(CEGB)}(r)= 1+\frac{r^2}{2\alpha} \left[ 1-\sqrt{1+4 \alpha \left( \frac{2M}{r^3}-\frac{Q^2}{r^4} \right)} \right].
\label{RNEGB}
\end{equation}
This  solution was first introduced in \cite{Fernandes2020}, in addition to a coupling with anti-de Sitter space. The asymptotic behavior of this solution is
\begin{eqnarray}
f^{(CEGB)}(r)& = & 1-\frac{2M}{r}+ \frac{Q^2}{r^2}+ \frac{4 M^2 \alpha}{r^4}\\ &&- \frac{4 M Q^2 \alpha}{r^5}+\mathcal{O}\left(\frac{1}{r^{6}}\right). \notag
\label{RNEGBapp}
\end{eqnarray}

When only the first and second orders are considered in the series expansion, the solutions of GR are clearly revealed.

For simplicity, from now on, the spacetime that describes the background geometry of an electrically charged BH in EGB 4D gravity of equation (\ref{RNEGB}) will be denoted as the CEGB BH solution.

\begin{figure*}
    \centering
\includegraphics[width=1\linewidth]{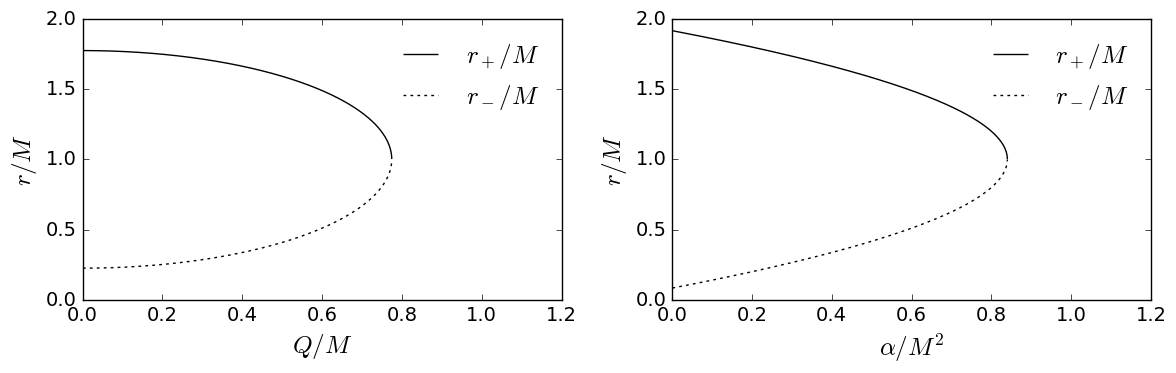}
    \caption{Radius of the event horizon $r_{+}$ and of the inner horizon $r_{-}$ for a CEGB BH. In the left panel it is shown in terms of the electric charge $Q$ (using $\alpha=0.1$) and in the right panel in terms of the GB coupling constant $\alpha$ (with $Q=0.1$).}
    \label{fig:fig1}
\end{figure*}

\subsection{Horizons and Particular Cases}

The CEGB BH solution has two horizons, the internal (Cauchy) horizon, $r_{-}$, and the event horizon, $r_{+}$, which are located at
\begin{equation}
r_{\pm}=M \pm \sqrt{M^{2}-Q^{2}-\alpha}.
\label{2-75}
\end{equation}
The corresponding behavior of $r_{\pm}$ is shown in Figure (\ref{fig:fig1}) for different values of $Q$ and $\alpha$. There, the monotonic corrections on $r_{\pm}$ are evident, at higher values of $\alpha$ and $Q$, the value of $r_{+}$ decreases while $r_{-}$ increases. Thus, it is possible to obtain a suitable value of the parameters $\alpha$ and $Q$ for which $r_{-}$ and $r_{+}$ become a single degenerate horizon (with $r_{-}= r_{+}=M$). This particular case is known as the extreme BH type and it is obtained when
\begin{equation}
M=M_{ext}=\sqrt{Q^{2}+\alpha}.
\end{equation}
Thus, when $M=M_{ext}$, the function $f^{(CEGB)}$ has only one real root (corresponding to the degenerate horizon). When $M>M_{ext}$ the usual BHs solutions are obtained, with two real roots representing the two horizon as in  equation (\ref{2-75}). Taking $M=1$ and  $M>M_{ext}$, $\alpha$ takes values in the range
\begin{equation}
  0< \alpha \leq 1-Q^2.
  \label{alpharange}
\end{equation}
From here it is clear that when $Q \rightarrow 1$, the parameter $\alpha \rightarrow 0$. On the other hand, when $M<M_{ext}$ we have two complex root and the solution will represent a naked singularity.

The BH solution $f^{(CEGB)}$, contains various types of BHs as particular limiting cases. First, the Schwarzschild space-time is reached when $Q \rightarrow 0$ and $\alpha \rightarrow 0$, obtaining
     \begin{equation}
f^{(Sch)}= 1-\frac{2M}{r} .
\label{Sch}
\end{equation}

The  Reissner–Nordström (RN) BH is achieved by taking $\alpha \rightarrow 0$ in the solution $f^{(CEGB)}$,
     \begin{equation}
f^{(RN)}= 1-\frac{2M}{r}+\frac{Q^2}{r^2} .
\label{RN}
\end{equation}
This metric describes the electrically charged BHs of the theory of GR and the horizons are given by
\begin{equation}
r_{\pm}=M \pm \sqrt{M^{2}-Q^{2}}.
\end{equation}
The third particular limiting case is the BH of EGB 4D, obtained when $Q \rightarrow 0$,
     \begin{equation}
f^{(EGB)}= 1+\frac{r^2}{2\alpha} \left( 1 - \sqrt{1+\frac{8M \alpha}{r^3} } \right).
\label{EGB}
\end{equation}
This spacetime has two real roots given by
\begin{equation}
r_{\pm}=M \pm \sqrt{M^{2}-\alpha}.
\end{equation} 

\section{Quasinormal Modes of the Test Fields Perturbations}

\subsection{Master Wave Equations of Scalar and Electromagnetic Fields}

A massless scalar field, $\Phi$, is described by the Klein-Gordon equation that  is written, using the background  metric $g_{\mu \nu}$, as 
\begin{equation}
\frac{1}{\sqrt{-g}} \partial_\mu \left( \sqrt{-g} g^{\mu \nu} \partial_\nu \Phi  \right)   =0 . \label{Mastereq1}
\end{equation}

On the other hand, the electromagnetic field equation in a curved spacetime is
\begin{equation}
\frac{1}{\sqrt{-g}} \partial_\mu \left( \sqrt{-g} F^{\mu \nu}  \right)   =0 . \label{Mastereq2}
\end{equation}

Using the formalism of perturbation theory and the scalar and vector harmonics to separate the spherical coordinates ($t,r, \theta, \phi$), equations \eqref{Mastereq1} and \eqref{Mastereq2} can be transformed into a single general differential equation that adopts a Schrödinger-like form for stationary backgrounds \cite{Konoplya2011},
\begin{equation}
\frac{d^2 \Psi_s}{dr_*^2}  + \left[ \omega^2 - V_s(r)  \right] \Psi_s = 0 , \label{ecuacionmaestrageneral}
\end{equation}
where we have introduced the well-known tortoise coordinate, $r_*$, defined by the relation
\begin{equation}
dr_* = \frac{dr}{f(r)}\label{Coordenadatortugageneral}
\end{equation}
and the effective potential takes de generalized form
\begin{equation}
V_s (r) = f(r) \left(  \frac{\ell(\ell+1)}{r^2} + \frac{(1-s)}{r} f '(r) \right). \label{potencialRWgeneral}
\end{equation}
In this  expression, $s=0$ and $s=1$ identify the scalar and the electromagnetic perturbations, respectively. Also, the prime denotes differentiation with respect to $r$, and $\ell=0,1,2, \dots$ are the multipole quantum numbers that come from spherical harmonic expansions. 

The QNMs frequencies, denoted by $\omega$ in equation \eqref{ecuacionmaestrageneral}, are obtained by requiring purely outgoing waves at infinity and purely incoming waves at the event horizon \cite{Konoplya2020},

\begin{equation}
\Psi_s \sim \pm e^{\pm i \omega r_*}, \quad r_* \rightarrow \pm \infty.
\end{equation}

In the case of the CEGB BH, the effective potential is always positive and have the shape of a potential barrier with a single peak. It also fulfills that
\begin{equation}
V_s^{(CEGB)}\left(r \rightarrow r_{+}\right)=V_s^{(CEGB)}(r \rightarrow \infty)=0 .
\end{equation}

This behavior of the effective potential are the necessary boundary conditions to use  the semi-analytical methods in the upcoming sections to calculate the QNMs frequencies.

\subsection{The WKB Aproximation Method}

Among the first theoretical approaches developed to calculate these  QNMs frequencies in a semi-analytical manner is the well known Wentzel-Kramers-Brillouin (WKB) approximation method \cite{wkb1}. The derivation presented by Schutz and Will \cite{wkb1} begins with a series expansion of the redefined potential $\Lambda(r_{*})=\omega^2 - V_s(r_{*})$. The value of the turtle coordinate at which the maximum point of the effective potential is reached will be denoted by $\tilde{r}_{*}$ and the potential evaluated at this point would be $V_s(\tilde{r}_{*}) =V_0$. Hence, the series expansion will be 
\begin{equation}
\Lambda=\Lambda_0+\frac{1}{2} \Lambda_0^{\prime \prime}\left(r_{*}-\tilde{r}_{*}\right)^2+\mathcal{O}\left(r_{*}-\tilde{r}_{*}\right)^3+...
\end{equation}
with $\Lambda_0=\omega^2-V_0 $ and  $\Lambda_0^{\prime \prime}=-V_0^{\prime \prime}$.
It is clear that the second term of the expansion corresponds to the condition of the maximum point of the potential and therefore it vanishes. Substituting this expansion into the differential equation \eqref{ecuacionmaestrageneral}, the master wave equation reduces to the parabolic cylinder differential equation (usually called the Weber equation), which has known solutions. Using the asymptotic behavior described above and imposing the boundary conditions that represent a BH, it is possible to find a simple analytical expression of the frequencies of the QNMs. At first order, it takes the form
\begin{equation}
\omega^2=V_0-i\sqrt{-2 V_0^{\prime \prime}}\left(n+\frac{1}{2}\right).
\label{203}
\end{equation}
where the expression is labeled with the harmonic or overtone number $n$. Both, the real part $\omega_R$ and the imaginary part $\omega_I$, as well as the overtone number $n$, depend only on the maximum potential, $V_0$, and on the second derivative of the potential evaluated in the maximum point, $V_0^{\prime \prime}$. It should be noted that although the WKB formula has been derived analytically, it is not always possible to find the value of $\tilde{r}_{*}$ explicitly. Therefore, it could be said that the WKB approach is a semi-analytic methodology. When this method was introduced, the QNMs of the gravitational perturbations of the Schwarzschild BH were estimated with an error of approximately 6\% \cite{Konoplya2019,wkb1}.

In 1987, Iyer and Will \cite{sai1987} extended the WKB approximation method up to the $3^{rd}$ order, improving the precision of the method up to an estimated error of less than 1\% for $n=0$ \cite{Konoplya2019}. The formula for the frequencies of the QNMs of the $3^{rd}$ order of the WKB method is
\cite{sai1987}
\begin{equation}
\omega^2=\left[V_0+\sqrt{-2 V_0^{\prime \prime}} \tilde{\Gamma}_1\right]-i\tilde{\Lambda}\sqrt{-2 V_0^{\prime \prime}}[1+\tilde{\Gamma}_2],
\label{204}
\end{equation}
where
\begin{widetext}
\begin{subequations}
\begin{align}
\tilde{\Gamma}_1&=\frac{1}{\sqrt{-2 V_0^{\prime \prime}}}\left[\frac{1}{8}\left(\frac{V_0^{(4)}}{V_0^{\prime \prime}}\right)\left(\frac{1}{4}+\tilde{\Lambda}^2\right)-\frac{1}{288}\left(\frac{V_0^{\prime \prime \prime}}{V_0^{\prime \prime}}\right)^2\left(7+60 \tilde{\Lambda}^2\right)\right],\label{3order1}\\
\tilde{\Gamma}_2 &=-\frac{1}{2 V_0^{\prime \prime}}\left[\frac{5}{6912}\left(\frac{V_0^{\prime \prime \prime}}{V_0^{\prime \prime}}\right)^4\left(77+188 \tilde{\Lambda}^2\right)-\frac{1}{384} \frac{V_0^{\prime \prime \prime 2} V_0^{(4)}}{V_0^{\prime \prime 3}}\left(51+100 \tilde{\Lambda}^2\right)\right.\label{3order2} \\ 
&+\frac{1}{2304}\left(\frac{V_0^{(4)}}{V_0^{\prime \prime}}\right)^2\left(67+68 \tilde{\Lambda}^2\right)+\frac{1}{288} \frac{V_0^{\prime \prime \prime} V_0^{(5)}}{V_0^{\prime \prime 2}}\left(19+28 \tilde{\Lambda}^2\right)\left.-\frac{1}{288} \frac{V_0^{(6)}}{V_0^{\prime \prime}}\left(5+4 \tilde{\Lambda}^2\right)\right] \notag ,
\end{align}
\end{subequations}
\end{widetext}
with $\tilde{\Lambda}=n+1/2$ and  $V_0^{\prime \prime \prime}$ and $V_0^{(j)}$ the third and $j^{th}$ derivatives of the potential, respectively, evaluated at the radial coordinate of the maximum point. 

In 2003, Konoplya \cite{Konoplya2003} extended the  WKB method to $6^{th}$, giving more accurate results than the previous expressions \cite{Konoplya2019}. In this case, the frequencies are given by the relation 
\begin{equation}
\frac{i\left(\omega^2-V_0\right)}{\sqrt{-2 V_0^{\prime \prime}}}-\sum_{j=2}^6 \Gamma_j=n+\frac{ 1}{2},
\label{WKB6}
\end{equation}
where the high order contributions, represented by  $\Gamma_j$, are not reproduced here because of their extension but can be consulted in reference \cite{Konoplya2003}. This equation depends on terms up to $V_0^{(12)}$, that is, the twelfth derivative of the potential evaluated at the radial coordinate of the maximum point. 

Finally, in 2017, Matyjasek and Opala \cite{matyjasek2017} developed the extension of the WKB method up to $13^{th}$ order. However, it has been shown that convergence in each order is not guaranteed and that the inclusion of more orders in the expansion does not ensure more accurate results \cite{Konoplya2019}. In any case, the equations of the WKB method at $1^{st}$, $3^{rd}$ and $6^{th}$ order give satisfactory results as long as $\ell>n$, with the best results obtained when $\ell \gg n$ and acceptable results when $\ell=n$ \cite{Konoplya2019}.

\subsection{The Pöschl–Teller Potential Method}

Another of the semi-analytical formalisms developed to calculate the frequencies of the QNMs of a BH was introduced in 1984 by Ferrari and Mashhoon \cite{Ferrari1984}. It consists in an approximation the effective potential included in the equation \eqref{potencialRWgeneral}  to the well known Pöschl-Teller potential, such that the master wave equation could be rewritten as
\begin{equation}
\frac{\partial^2 \Psi}{\partial r_*^2}+\left[\omega^2-\frac{V_0}{\cosh ^2 \eta\left(r_*-\bar{r}_*\right)}\right] \Psi=0,
\label{207}
\end{equation}
with
\begin{equation}
\eta^2=\frac{V_0^{\prime \prime}}{2 V_0}.
\end{equation}

Making some substitutions and following a similar procedure as that in the WKB approximation, it is possible to transform  equation \eqref{207} into a differential equation whose solutions are the hypergeometric functions \cite{Berti2009}. Analyzing their asymptotic behavior, it is possible to develop a semi-analytic formula for the frequencies of QNMs,
\begin{equation}
\omega=\pm \sqrt{V_0-\frac{\eta^2}{4}}-i\eta( n+\frac{1}{2}) .
\label{PT}
\end{equation}

This approach considers both the real and imaginary components of the frequency to be dependent on the potential and its second derivative evaluated at the point of maximum. However, only the imaginary component, $\omega_I$, is affected by the overtone number, $n$. This method can provide more accurate results for $\omega_I$ compared to those obtained using the WKB formula at $1^{st}$ order of the equation \eqref{203}. Hence, this treatment is not recommended for determining the real component $\omega_R$ of the perturbation frequencies, except in specific cases such as the eikonal limit ($\ell \rightarrow \infty$) or for the fundamental mode (n=0) \cite{Berti2009}. In order to ensure greater precision in our discussions, in the upcoming sections we will keep these restrictions in mind and primarily focus on the eikonal QNMs of the fundamental mode (n=0).\\

In general, these semi-analytical formulas do not provide precise results when $n \geq \ell$ or when the potential contains divergences, as is the case of perturbations of some massive scalar fields or for asymptotically deSitter and Anti-deSitter spaces. In these situations, the conditions required by the formalisms are not met, since they require the ability to identify the characteristic maximum point of the potential barrier.  Consequently, other alternative approaches have been proposed for calculating QNM frequencies, including classical and numerical methods such as the Chandrasekhar-Detweiler method, direct integration of the wave equation, the Frobenius series method and its variations, the continued fractions method, and the monodromy technique for highly damped QNMs (for a discussion of these methods see \cite{Konoplya2011, Berti2009}). In recent years, novel and alternative computational methods have also been developed to obtain BH QNM frequencies, such as the Borel summation method \cite{Hatsuda2020}, the Jansen Mathematica package \cite{Jansen} or  the use of Neural Networks Methods \cite{networks}.

\section{Quasinormal Modes in the Eikonal Regime}

\begin{figure*}
    \centering
\includegraphics[width=1\linewidth]{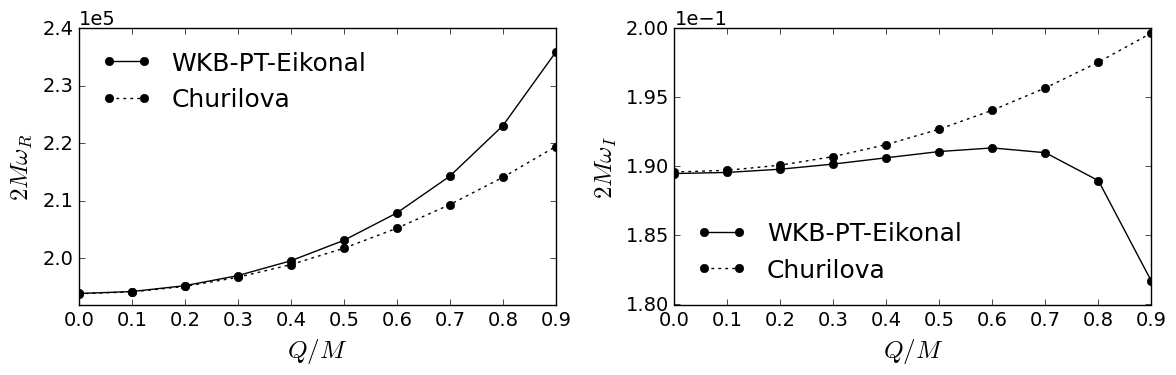}
    \caption{Eikonal QNMs frecuencies for a CEGB BH depending on the electric charge $Q$. The left 
 panel corresponds to the behavior of the real part of the frequencies while the right panel illustrates the behavior of the imaginary part. (using $n=0$, $\ell=500000$ and $\alpha=0.1$).}
    \label{fig:fig2}
\end{figure*}

\begin{figure*}
    \centering
\includegraphics[width=1\linewidth]{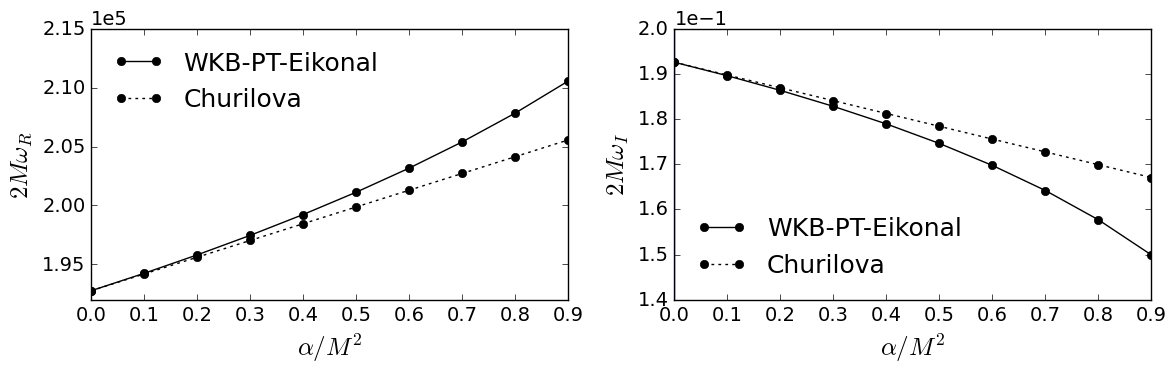}
    \caption{Eikonal QNMs frecuencies for a CEGB BH depending on the GB coupling constant $\alpha$. The left panel corresponds to the behavior of the real part of the frequencies while the right panel illustrates the behavior of the imaginary part. (using $n=0$, $\ell=500000$ and $Q=0.1$).}
    \label{fig:fig3}
\end{figure*}

\subsection{Eikonal QNMs Approaches}

It has been established in \cite{Cardoso2009} that there is a strong correlation between null geodesics and QNMs. In particular, in the eikonal regime, the real and imaginary parts of QNMs frequencies for any spherically symmetric, asymptotically flat spacetime can be related to the frequency and instability time scale of unstable circular null geodesics. Therefore, in the eikonal limit, we can establish a correlation between the radius of the photon sphere $R_{ps}$ of a BH and its QNMs using the following analytical expression  \cite{Jusufi2020, Chen2021, Mishra2020}
\begin{equation}
\omega=\Omega \ell-i\left(n+\frac{1}{2}\right)\lambda .
\label{eikonalomega}
\end{equation}
where $\Omega$ is the angular velocity at the photon sphere,
\begin{equation}
\Omega=\sqrt{\frac{f^{\prime}(R_{ps})}{2 R_{ps}}}.
\label{eikonalomega2}
\end{equation}

The symbol $\lambda$ in \eqref{eikonalomega} represents the Lyapunov exponent, which can be expressed as
\begin{equation}
\lambda=\sqrt{\frac{f\left(R_{ps}\right)\left(2 f\left(R_{ps}\right)-R_{ps}^2 f^{\prime \prime}\left(R_{ps}\right)\right.}{2 R_{ps}^2}} .
\label{eikonalomega3}
\end{equation}

This parameter is associated with the instability time scale of the photon sphere of a BH, indicating how quickly the orbit becomes unstable.

An interesting approach to the eikonal limit was proposed  by M. Churilova \cite{Churilova2019} by noting that  the effective potential $V_{eik}$ in this limit does not usually depend on the spin of the field, except for some exceptions such as the backgrounds of charged BHs coupled to non-linear electromagnetic fields or the gravitational perturbations in some theories with higher curvature corrections, like EGB, Einstein–Lovelock or Einstein–dilaton–Gauss–Bonnet theories. Therefore, in most static and spherically symmetric spactimes, the effective potential in the eikonal aproximation  for scalar and electromagnetic perturbations can be expressed as \cite{Churilova2019}
\begin{equation}
V_{eik}(r)=f(r)\left(\frac{\ell(\ell+1)}{r^2}+\mathcal{O}(1)\right).
\label{eikonalpotential}
\end{equation}

This means that the effective potential $V_{eik}$ of the  test fields perturbations in the eikonal limit  can have the same form as the potential of the electromagnetic field perturbations $V_{s=1}$. We will analyze this fact below on the CEGB BH, with the help of our results of the QNMs frequencies of the scalar and electromagnetic field using high values of $\ell$. 
 
Additionally, in \cite{Churilova2019} a general approach for eikonal QNMs of asymptotically flat BH solutions is presented. There, an expansion of the first order WKB formula of the equation \eqref{203} is written in powers of small parameters defined by the deviations of a given metric from the Schwarzschild one. These small parameters on the CEGB BH can be identified in the metric expansion of equation \eqref{RNEGBapp}. Consequently, applying this Churilova analytical formula, in the eikonal limit, for a CEGB BH we obtain 
\begin{widetext}
\begin{equation}
\begin{aligned}
\omega^{(Ch)}= & \frac{\left(\ell+\frac{1}{2}\right)}{3 \sqrt{3} M}\left(1+\frac{Q^2}{6 M^2}+\frac{2 \alpha}{27 M^2}-\frac{2 Q^2\alpha}{81 M^4}\right) -i \frac{\left(n+\frac{1}{2}\right)}{3 \sqrt{3} M}\left(1+\frac{Q^2}{18 M^2}-\frac{ 4\alpha}{27 M^2}+\frac{22 Q^2 \alpha}{243 M^4}\right) +\mathcal{O}\left(\frac{1}{\ell+\frac{1}{2}}\right)
\label{ChurilovaOmega}
\end{aligned}
\end{equation}
\end{widetext}

This result reproduces the analytical form of the eikonal QNMs of the EGB 4D BH found in \cite{Konoplya2020}, when $Q=0$. Using equations \eqref{eikonalomega} and \eqref{ChurilovaOmega}  we have that the angular velocity of the photon sphere is approximately
\begin{equation}
\Omega^{(Ch)}=\frac{(2 \ell+1) \left(162 M^4+3 M^2 \left(4 \alpha +9 Q^2\right)-4 \alpha  Q^2\right)}{972 \sqrt{3} \ell M^5},
\end{equation}
and the Lyapunov exponent takes the form
\begin{equation}
\lambda^{(Ch)}=\frac{486 M^4+9 M^2 \left(3 Q^2-8 \alpha \right)+44 \alpha  Q^2}{1458 \sqrt{3} M^5}.
\end{equation}

In the following sections, we will obtian the radius of the shadow of a CEGB BH from $\omega^{(Ch)}$. This approximate expression will not only depend on the geometric parameters of the BH solution, but also on $\ell$, so their applicability will be limited to the eikonal regime.

\subsection{Eikonal QNMs of a CEGB BH}

\begin{table*}
\caption{\label{tab:table1}QNMs frequencies of the scalar and electromagnetic fields around of a CEGB BH and for various values of $n$. (with $\ell=500000$, $\alpha=0.1$ and $Q=0.1$).}
\begin{ruledtabular}
\begin{tabular}{cccc}
&\multicolumn{1}{c}{}&\multicolumn{1}{c}{$2 M \omega$}  \\ \hline
\vspace{-0.3cm}\\
 $n$ &  WKB $1^{st}$, $3^{rd}$ and $6^{th}$ order and PT & Eikonal & Churilova \vspace{0.1cm}\\ \hline
 0 & 194248.7496-0.1896i & 194248.5553-0.1896i & 194191.8384-0.1897i \\

 1 & 194248.7496-0.5687i & 194248.5553-0.5687i & 194191.8384-0.5692i  \\

 2 & 194248.7496-0.9478i & 194248.5553-0.9478i & 194191.8384-0.9486i  \\

 3 & 194248.7496-1.3270i & 194248.5553-1.3270i & 194191.8384-1.3281i  \\

 4 & 194248.7496-1.7061i & 194248.5553-1.7061i & 194191.8384-1.7075i  \\

 5 & 194248.7496-2.0853i & 194248.5553-2.0853i & 194191.8384-2.0870i  \\

 6 & 194248.7496-2.4644i & 194248.5553-2.4644i & 194191.8384-2.4664i  \\

 7 & 194248.7496-2.8435i & 194248.5553-2.8435i & 194191.8384-2.8458i  \\

 8 & 194248.7496-3.2227i & 194248.5553-3.2227i & 194191.8384-3.2253i  \\

  9 & 194248.7496-3.6018i & 194248.5553-3.6018i & 194191.8384-3.6047i  \\
\end{tabular}
\end{ruledtabular}
\end{table*}

 To calculate and study the eikonal QNMs of the CEGB BH, we use the WKB aproximation formulas at $1^{st}$, $3^{rd}$, and $6^{th}$ order, given by equations \eqref{203}, \eqref{204}, and \eqref{WKB6}, respectively. We also calculate the eikonal frequencies of the QNMs through the PT potential using equation \eqref{PT}, the eikonal formula given by the equation \eqref{eikonalomega}, and the Churilova's analytical formula in equation \eqref{ChurilovaOmega}. We show some of these results in Table \ref{tab:table2} for the QNMs of scalar perturbations (s=0), in Table \ref{tab:table3} for the QNMs of electromagnetic perturbations (s=1) and in Table \ref{tab:table1} for both test fields.

In Table \ref{tab:table1}, we present the QNMs frequencies of the scalar and electromagnetic fields around a CEGB BH for $\ell = 500000$ and various values of $n$. The second column summarize the results obtained from the WKB method in three diferent orders and the PT method, which gave the same results. The third and fourth columns list the values calculated using the eikonal and Churilova formulas, respectively. It is evident that the real part of the frequencies does not depend on $n$ in the eikonal regime. Additionally, it can be seen that the imaginary part of the frequencies obtained from the WKB and PT methods closely match the results calculated using the eikonal limit formula. Although Churilova's eikonal formula is not as close to the other results, it still agrees well with them. Considering these facts, as depicted in Figures (\ref{fig:fig2}) and (\ref{fig:fig3}), the curves representing the frequency behavior overlap in the solid line that summarizes the behavior of the WKB, PT and the eikonal methods, but differs from the dotted line that displays the results of the Churilova formula. 

In Tables \ref{tab:table2} and \ref{tab:table3} we note that  increasing $\ell$ implies that these methods consistently yield similar values. It is also clear that the imaginary part of the frequencies does not depend on $\ell$, as is foreseen by the analytical expressions of the eikonal limit and Churilova formulas in equations \eqref{eikonalomega} and \eqref{ChurilovaOmega}, respectively.\\

For a CEGB BH, exactly the same values of the frequencies $\omega$ are obtained in all the methods when $\ell>50000$ for the perturbations of both the scalar and the electromagnetic field. For the eikonal and Churilova formulas, this is an obvious result because they do not depend on the field. However, this result for the WKB and PT methods proves the convergence of the eikonal QNMs frequencies. Therefore, the effective potential of these test fields over the CEGB BH effectively behaves like equation \eqref{eikonalpotential} in the eikonal limit.

Similarly, our results for the WKB and PT methods when $\ell < 50000$ show that the real and imaginary parts of the frequencies $\omega$ are, in general, a little smaller for the electromagnetic field than for the scalar field, but  they converge for $\ell \approx 50000$, as illustrated in  Table \ref{tab:table1}.

\begin{table*}
\caption{\label{tab:table2}QNMs frequencies of the scalar field (s=0) around of a CEGB BH and for various high values of $\ell$. (with $n=0$, $\alpha=0.1$ and $Q=0.1$).}
\begin{ruledtabular}
\begin{tabular}{ccccccc}
&\multicolumn{2}{c}{}&\multicolumn{2}{c}{$2 M \omega$} \\
\hline
\vspace{-0.3cm}\\
 $\ell$ & WKB $1^{st}$ order & WKB $3^{rd}$ order & WKB $6^{th}$ order & PT & Eikonal & Churilova \vspace{0.1cm}\\ \hline
 5 & 2.1594-0.1895i & 2.1390-0.1898i & 2.1391-0.1898i & 2.1426-0.1903i & 1.9425-0.1896i & 2.1361-0.1897i \\

 10 & 4.0911-0.1896i & 4.0804-0.1896i & 4.0805-0.1896i & 4.0823-0.1898i & 3.8850-0.1896i & 4.0780-0.1897i \\

 50 & 19.6216-0.1896i & 19.6194-0.1896i & 19.6194-0.1896i & 19.6197-0.1896i & 19.4249-0.1896i & 19.6134-0.1897i \\

 100 & 39.0452-0.1896i & 39.0441-0.1896i & 39.0441-0.1896i & 39.0443-0.1896i & 38.8497-0.1896i & 39.0325-0.1897i \\

 500 & 194.4431-0.1896i & 194.4428-0.1896i & 194.4428-0.1896i & 194.4429-0.1896i & 194.2486-0.1896i & 194.3858-0.1897i \\

 1000 & 388.6915-0.1896i & 388.6914-0.1896i & 388.6914-0.1896i & 388.6914-0.1896i & 388.4971-0.1896i & 388.5775-0.1897i \\

 5000 & 1942.6798-0.1896i & 1942.6798-0.1896i & 1942.6798-0.1896i & 1942.6798-0.1896i & 1942.4856-0.1896i & 1942.1106-0.1897i \\

 10000 & 3885.1654-0.1896i & 3885.1654-0.1896i & 3885.1654-0.1896i & 3885.1654-0.1896i & 3884.9711-0.1896i & 3884.0271-0.1897i \\

 50000 & 19425.0498-0.1896i & 19425.0498-0.1896i & 19425.0498-0.1896i & 19425.0498-0.1896i & 19424.8555-0.1896i & 19419.3586-0.1897i \\
\end{tabular}
\end{ruledtabular}
\end{table*}

\begin{table*}
\caption{\label{tab:table3}QNMs frequencies of the electromagnetic field (s=1) around of a CEGB BH and for various high values of $\ell$. (with $n=0$, $\alpha=0.2$ and $Q=0.2$).}
\begin{ruledtabular}
\begin{tabular}{ccccccc}
&\multicolumn{2}{c}{}&\multicolumn{2}{c}{$2 M \omega$}\\ \hline
\vspace{-0.3cm}\\
 $\ell$ & WKB $1^{st}$ order & WKB $3^{rd}$ order & WKB $6^{th}$ order & PT & Eikonal & Churilova \vspace{0.1cm}\\ \hline
 5 & 2.1646-0.1858i & 2.1450-0.1859i & 2.1451-0.1859i & 2.1486-0.1865i & 1.9687-0.1865i & 2.1620-0.1873i \\

 10 & 4.1338-0.1863i & 4.1236-0.1863i & 4.1236-0.1863i & 4.1254-0.1865i & 3.9374-0.1865i & 4.1275-0.1873i \\

 50 & 19.8840-0.1865i & 19.8819-0.1865i & 19.8819-0.1865i & 19.8822-0.1865i & 19.6872-0.1865i & 19.8512-0.1873i \\

 100 & 39.5713-0.1865i & 39.5702-0.1865i & 39.5702-0.1865i & 39.5704-0.1865i & 39.3744-0.1865i & 39.5058-0.1873i \\

 500 & 197.0691-0.1865i & 197.0688-0.1865i & 197.0688-0.1865i & 197.0689-0.1865i & 196.8722-0.1865i & 196.7427-0.1873i \\

 1000 & 393.9413-0.1865i & 393.9411-0.1865i & 393.9411-0.1865i & 393.9412-0.1865i & 393.7444-0.1865i & 393.2889-0.1873i \\

 5000 & 1968.9188-0.1865i & 1968.9188-0.1865i & 1968.9188-0.1865i & 1968.9188-0.1865i & 1968.7219-0.1865i & 1965.6584-0.1873i \\

 10000 & 3937.6407-0.1865i & 3937.6407-0.1865i & 3937.6407-0.1865i & 3937.6407-0.1865i & 3937.4439-0.1865i & 3931.1203-0.1873i \\

 50000 & 19687.4161-0.1865i & 19687.4161-0.1865i & 19687.4161-0.1865i & 19687.4161-0.1865i & 19687.2193-0.1865i & 19654.8153-0.1873i \\
\end{tabular}
\end{ruledtabular}
\end{table*}

Taking $\omega=\omega_R-\omega_I$, in Figures (\ref{fig:fig2}) and (\ref{fig:fig3}),  we show the effect of the geometric parameters $Q$ and $\alpha$ on the real and imaginary components of the eikonal QNMs frequencies for a CEGB BH.  Our results show that, as the parameter $\alpha$ increases, the real and imaginary part of the QNMs frequencies $\omega_{R}$ and $\omega_{I}$ increase monotonically as well. On the other hand, when the electric charge $Q$ grows, the real part $\omega_{R}$ also increases but the imaginary part of the frequencies increases up to a maximum peak of growth to decrease after it is reached. We obtained the same corrections due to $Q$ and $\alpha$ on the QNMs of the CEGB BH that were reported in \cite{Chen2021}. However, our results show that when $Q$ increases, $\omega_{I}$ also increases, but for values close to $Q^2=M^2-\alpha$, as shown in the figure (\ref{fig:fig2}), $\omega_{I}$, it begins to decrease. In reference \cite{Ferrari1984}, where the PT potential method was introduced, these same behaviors were found due to the electric charge on the QNMs of the RN BH. Additionally, in \cite{Panotopoulos2018}, a similar effect of the electric charge was found on the frequencies of the QNMs of a charged BH in Einstein-Maxwell non-linear electrodynamics. For the CEGB BH, these corrections on the QNMs caused by the electric charge are consistent with those recently reported in \cite{Zhang2020}. Furthermore, these effects of the GB coupling constant $\alpha$ on the real and imaginary part of the QNMs of the CEGB BH agree with the results reported in 
\cite{Zhang2020, Konoplya2020, Chen2021, Liu2022, Devi2020}.

Another interesting observation is that the QNMs frequencies calculated by the Churilova formula present a better agreement with other results in the literature when the geometric parameters of the BH are small. This is expected, given the nature of the solution's expansion \cite{Churilova2019}.  Therefore, the Churilova formula's frequencies differ from other results as $M$ approaches $M_{ext}$.

\section{The effective Potential and the Photon Sphere}

\begin{figure*}
    \centering
\includegraphics[width=1\linewidth]{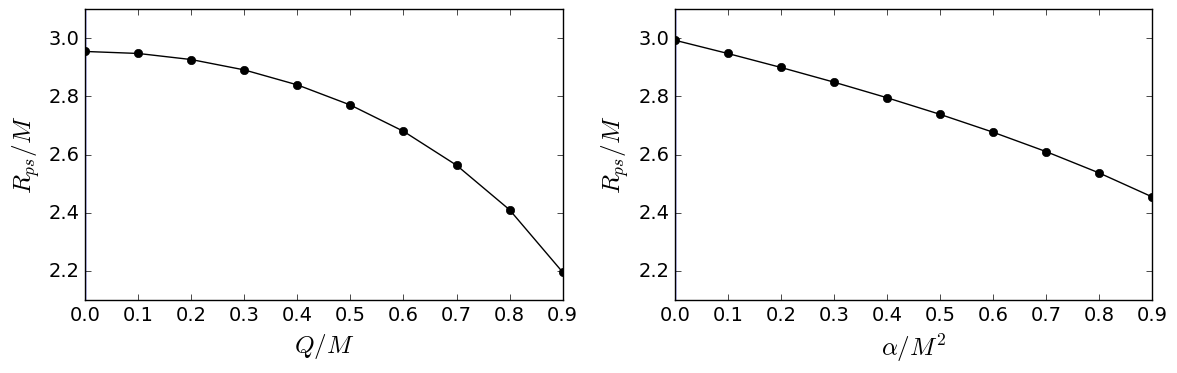}
    \caption{Photon sphere radius $R_{ps}$ for a CEGB BH using the equations (\ref{eq:RpsCEGB}). In the left panel it is shown in terms of the electric charge $Q$ (using $\alpha=0.1$) and in the right panel in terms of the GB coupling constant $\alpha$ (with $Q=0.1$).}
    \label{fig:fig4}
\end{figure*}

\subsection{The BH Photon Sphere Radius}

The photon sphere, also known as the "light ring" or "photon orbit", for a static spherically symmetric BH is recognized as the orbit at which light moves in a unstable circular null geodesic. As mentioned in \cite{Jusufi2020, Chen2021}, the Hamilton-Jacobi or Hamiltonian formulations can be used to find the equations of motion for photons around a static and spherically symmetric BH background and subsequently, permit to identify the effective potential that describes the system. From the critical point conditions of this potential, the radius of the photon sphere $R_{ps}$ can be determined  by solving the expression
\begin{equation}
2-   \frac{R_{ps} f^{\prime}(R_{ps})}{f(R_{ps})}=0 .
\label{photonsphere}
\end{equation}

Substituting the Schwarzschild metric on this expression, we  obtain that
\begin{equation}
R_{ps}^{(Sch)}=3 M.
\label{eq:RpsSch}
\end{equation}

Using the solution of RN given by \eqref{RN}, we have
\begin{equation}
R_{ps}^{(RN)}=\frac{1}{2} \left(\sqrt{9 M^2-8 Q^2}+3 M\right)
\label{eq:RpsRN}
\end{equation}

It should be noted that this  expression for $R_{ps}^{(RN)}$ has the same analytical form given in \cite{Konoplya2018} for the radial coordinate of the maximum  of the corresponding effective potential of field perturbations in the eikonal limit. Additionally,  the expression for $R_{ps}^{(RN)}$ can be approximated for small values of $Q$ as
\begin{equation}
R_{ps}^{(RN)}=3 M-\frac{2 Q^2}{3 M}-\frac{4 Q^4}{27 M^3}+O\left(Q^5\right).
\end{equation}

Using the EGB BH metric given by equation \eqref{EGB} in the condition of the equation \eqref{photonsphere}, we get
\begin{equation}
R_{ps}^{(EGB)}=\frac{\left(M \left(\sqrt{16 \alpha ^2-27 M^4}-4 \alpha \right)\right)^{2/3}+3 M^2}{\left(M \left(\sqrt{16
   \alpha ^2-27 M^4}-4 \alpha \right)\right)^{1/3}}.
   \label{eq:RpsEGB}
\end{equation}

In \cite{Guo2020}, another alternative analytical expression is deduced  for the radius of the photon sphere of the EGB 4D BH, with $M=1$,
\begin{equation}
R_{ps}^{(EGB)}=2 \sqrt{3} \cos \left(\frac{1}{3} \cos ^{-1}\left(-\frac{4 \alpha }{3 \sqrt{3}}\right)\right).
\end{equation}

 Although we couldn't establish an equality between these two expressions, we have tested them numerically and they always gave identical results. However, both expressions lead to the same expansion,
\begin{equation}
R_{ps}^{(EGB)}=3 M-\frac{4 \alpha }{9 M}-\frac{8 \alpha ^2}{81 M^3}+O\left(\alpha ^3\right),
\end{equation}
useful for small values of $\alpha$. Note that the expansions of $R_{ps}^{(RN)}$ and $R_{ps}^{(EGB)}$ show that the first term corresponds to radius of the photon spher for the Schwarzschild BH.

\subsection{The photon sphere radius for a CEGB BH}

Using the expression of equation \eqref{photonsphere}, an analytical expression for the radius of the photon sphere for a CEGB BH is obtained as

\begin{widetext}
\begin{subequations}
\begin{eqnarray}
R_{ps}^{(CEGB)}&=&\frac{1}{2} \sqrt{-\frac{8 M \left(2 \alpha +3 Q^2\right)}{\sqrt{\text{X}}}-\text{X}+18
   M^2}+\frac{\sqrt{\text{X}}}{2},\\ \text{X}&=&\frac{27 M^4-16 Q^2 \left(\alpha +Q^2\right)}{3^{1/3} \text{Y}}+\frac{\text{Y}}{3^{2/3}}+6 M^2
\\
\text{Y}&=&\left(-243 M^6+72 M^2 \left(4 \alpha ^2+3 Q^4+6 \alpha Q^2\right)+\text{Z}\right)^{1/3},\\
\text{Z}&=&\frac{1}{6} \sqrt{1728 \left(27 \alpha  M^4+4 Q^6\right) \left(64 \left(\alpha +Q^2\right)^3-27 M^4 \left(4 \alpha +3
   Q^2\right)\right)}
.
\end{eqnarray}
\label{eq:RpsCEGB}
\end{subequations}
\end{widetext}

The above equations for $R_{ps}^{(CEGB)}$ effectively reproduce the analytic expressions corresponding to the radius of the photon spheres of the solutions of the particular limit cases. Taking $Q \rightarrow 0$ and $\alpha \rightarrow 0$, we get $R_{ps}^{(Sch)}$ in equation \eqref{eq:RpsSch}. Doing only $\alpha \rightarrow 0$ gives $R_{ps}^{(RN)}$ given by equation \eqref{eq:RpsRN} and doing only $Q \rightarrow 0$ recovers $R_{ps}^{(EGB)}$ given by  equation \eqref{eq:RpsEGB}.

In figure (\ref{fig:fig4}),  we show the effect of the geometric parameters $Q$ and $\alpha$ on $R_{ps}^{(CEGB)}$. It is clear that as both the electric charge and the GB coupling constant increase, the photon sphere radius becomes smaller. These results agree with those reported in \cite{Guo2020, Panah2020}.

\subsection{Correspondence Between Eikonal QNMs and Null Geodesics}

As we astated above, there exist a correlation between null geodesics and eikonal QNMs frequencies for any spherically symmetric, asymptotically flat spacetime \cite{Cardoso2009}, which to a large extent can be linked to the fulfillment of the equation \eqref{eikonalomega}. However, with the simple fact that the BH solution is no longer valid for the WKB formula to first order, it is enough to believe that this connection does not hold. For example, the predicted correlation between null geodesics and QNMs is not upheld in the Einstein-Lovelock theory, as reported in \cite{Konoplya2017}, where authors showed that the radius of the photon sphere does not match the radial position of the extremum of the effective potential in the eikonal regime, contributing to the breakdown of the proposed correspondence.

In the previous section, we found that the effective potential of perturbations of the scalar and electromagnetic fields around the CEGB BH follows equation \eqref{eikonalpotential}  in the eikonal limit. However, this is not the case for gravitational perturbations in this background. In fact, in \cite{Konoplya2020}, it is shown that there is no correspondence between gravitational eikonal QNMs and null geodesics in the EGB 4D BH because of the form of the effective potential. They also mention that this correspondence does hold true for test fields perturbations when the background metric is considered to be a viable BH solution. Nevertheless, for scalar and electromagnetic fields in the CEGB BH spacetime, the QNMs calculations and its connections with the null geodesics and with the radius of the BH shadow should be satisfied and valid.

With the purpose of analyzing the correspondence in the CEGB BH, we will examine the agreement between the radius of the photon sphere, $R_{ps}^{(CEGB)}$, and the radial position of the maximum of the effective potential, $\tilde{r}$, for both scalar and electromagnetic field perturbations in the eikonal limit.

By differentiating with respect to the radial coordinate $r$ and equating to zero the equation \eqref{potencialRWgeneral} of the effective potential, $V_s (r)$, gives a condition that find the radial coordinate $\tilde{r}$ of the maximum point of the potential, as follows
\begin{widetext}
\begin{equation}
\frac{1}{r^3} \Big\{ rf' \left[  \ell(\ell+1)+r(1-s^2)f'  \right]-f\left[  2\ell(\ell+1)+r(1-s^2)\left( f'  -rf''\right)\right] \Big\}_{r=\tilde{r}} =0. \label{potencialmax}
\end{equation}
\end{widetext}

In Table \ref{tab:table4}, we present some values of the effective potential $V_s$ for scalar perturbations and the radial position of the maximum $\tilde{r}$ for various values of $\ell$, in a CEGB BH background. The results show that as $\ell$ increases, $V_s$ increases as well, but $\tilde{r}$ becomes closer to the correct value of the radius of the photon sphere that, for the values  $\alpha=0.1$ and $Q=0.1$, is $R_{ps}^{(CEGB)}=2.94761M$.
On the other hand, in the case of electromagnetic perturbations ($s=1$), the value of $\tilde{r}$ can be found analytically and it exactly matches the expression for $R_{ps}^{(CEGB)}$ given in \eqref{eq:RpsCEGB}. 
Consequently, this establishes that, in the eikonal limit, the CEGB BH also satisfies the connection between the maximum effective potential of the perturbations and the radius of the photon sphere, at least for scalar and electromagnetic fields.

\begin{table}[H]
\caption{\label{tab:table4}
The effective potential and its radial coordinate of the maximum $\tilde{r}$ for the scalar perturbations around a CEGB BH. We show the convergence for various values of $\ell$. (whit $\alpha=0.1$, $Q=0.1$ and $R_{ps}^{(CEGB)}=2.94761M$).}
\begin{ruledtabular}
\begin{tabular}{ccc}
\vspace{-0.3cm}\\
$\ell$ &$M^2 V_{s=0}$ &$\tilde{r}/M$\vspace{0.1cm}\\ \hline
5 & 1.15675 & 2.93786\\ 
10 & 4.17533 & 2.94489\\  
50 & 96.24260 & 2.94749\\  
100 & 381.12300 & 2.94758\\  
500 & 9452.01630 & 2.94761\\  
1000 & 37770.25849 & 2.94761\\  
5000 & 943501.21847 & 2.94761\\  
10000 & 3773627.47464 & 2.94761\\ 
50000 & 94333139.77196 & 2.94761\\  
\end{tabular}
\end{ruledtabular}
\end{table}

\section{The Black Hole Shadow}

\subsection{Connection Between Eikonal QNMs and the Radius of the BH Shadow}

The BH shadow is a region of darkness that results from the deflection of light by the strong gravitational field in the BH surroundings. Light from background objects that would normally reach an observer is instead absorbed by the BH, resulting in the appearance of a shadow. The shape and size of this shadow is a unique characteristic of each BH and it depends on its geometric properties as well as on the characteristics of the material surrounding it. 

Assuming a static observer positioned at a radial coordinate $r_O$, far enough from a spherically symmetric BH and such that $f(r_O) \approx 1$, the radius of the BH shadow, $R_{sh}$, as seen by this observer can be approximately calculated by \cite{Chen2021, Jusufi2020,Guo2020, Konoplya2020}
\begin{equation}
R_{sh} \approx \frac{R_{ps}}{\sqrt{f\left(R_{ps}\right)}}.
\label{Shadow1}
\end{equation}

It can be shown that for a Schwarzschild BH, the shadow radius is $R_{sh}^{(Sch)} = 3 \sqrt {3} M$. On the other hand, using the equation for $R_{ps}^{(CEGB)}$ in terms geometric parameters for the CEGB BH solution to obtain the shadow radius is not straightforward. Therefore, in order to find an expression for $R_{sh}^{(CEGB)}$, we will use the previous calculation of the QNMs as the starting point. Based on the correlations between the distance from the BH  and the behavior of eikonal QNMs with the photon sphere, it can be deduced that the real part of QNMs frequencies is inversely proportional to $R_{sh}$ \cite{Jusufi2020}. This relationship can be simply stated as follows
\begin{equation}
\omega_{R}=\lim _{\ell \gg 1} \frac{\ell}{R_{sh}}.
\label{Shadow2}
\end{equation}

In general terms, it is also proposed that the relationship between the QNMs and the shadow radius of spherically symmetrical BHs can be expressed as \cite{Chen2021}
\begin{equation}
\omega=R_{sh}^{-1}\left(\ell+\frac{D-3}{2}\right)-i\left(n+\frac{1}{2}\right) \lambda
\label{Shadow3}
\end{equation}
where $D$ represents the dimension of spacetime. In the eikonal limit, the term $(D-3)/2$ can be disregarded. Nevertheless, it can be useful to assess the connection between BH shadows and QNMs at small $\ell$, especially in spherically symmetrical spacetimes defined in high or lower dimensions. Moreover, a generalized equation linking eikonal QNMs and shadows of rotating BHs is provided in \cite{Jusufi2020-2}. Nonetheless, rotating BHs are not discussed in depth here but they hold significance for future related research projects.

\subsection{CEGB BH Shadow Radius}

\begin{figure*}
    \centering
\includegraphics[width=1\linewidth]{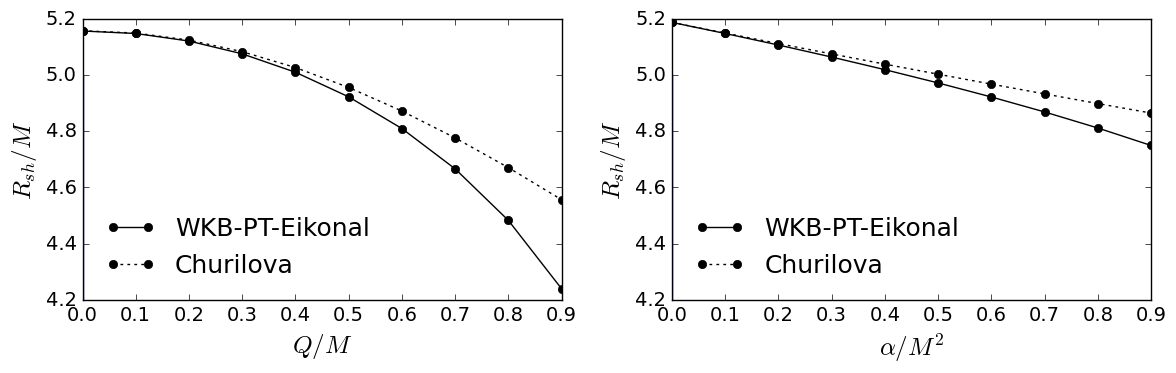}
    \caption{Shadow radius $R_{sh}$ for a CEGB BH (with $\ell=500000$). In the left panel it is shown in terms of the electric charge $Q$ (using $\alpha=0.1$) and in the right panel in terms of the GB coupling constant $\alpha$ (with $Q=0.1$). }
    \label{fig:fig5}
\end{figure*}

The connection between the shadow and the QNMs of the massless scalar field around a CEGB BH is explored in \cite{Chen2021}, but they do it using only the $6^{th}$ order WKB method and from the equation \eqref{Shadow3}. In this work,  we will use the equation \eqref{Shadow2}, which holds  only in the eikonal regime, and we will test several additional methods to study the shadow and QNMs of scalar and electromagnetic perturbations around a CEGB BH. 

To determine the shadow radius of the CEGB BH, we use the frequencies of the QNMs computed through the WKB method at $1^{st}$, $3^{rd}$, and $6^{th}$ order, as specified in equations \eqref{203}, \eqref{204}, and \eqref{WKB6}, respectively. We also  use the frequencies of the QNMs derived from the PT potential method outlined in equation \eqref{PT}, the eikonal formula that linking QNMs with $R_{ps}$ given by the equation \eqref{eikonalomega}, and the Churilova's analytical formula presented in equation \eqref{ChurilovaOmega}.

By substituting $R_{ps}^{(CEGB)}$ from the equations \eqref{eq:RpsCEGB} into the equation \eqref{Shadow1}, it is possible to provide a rather complicated analytical expression of the shadow radius of CEGB BH, which is not shared here because it is so extensive. However, from  Churilova's analytical eikonal approach, contrary to all the others studied, it is possible to provide an approximate analytical formula for the CEGB BH shadow radius. Using the equations \eqref{ChurilovaOmega} and \eqref{Shadow2}, it takes the form
\begin{equation}
    R_{sh}^{(Ch)}=\frac{972 \sqrt{3} l M^5}{(2 l+1) \left(162 M^4+3 M^2 \left(4 \alpha +9 Q^2\right)-4 \alpha  Q^2\right)}.
    \label{Shadow4}
\end{equation}

This CEGB BH shadow radius formula is related to the eikonal limit and therefore, in addition to the geometric parameters of the solution, it also depends on $\ell$. 

\begin{table}[H]
\caption{\label{tab:table5}
Shadow radius $R_{sh}$ for a CEGB BH using various values of $\ell$. (with $n=0$, $\alpha=0.1$, $Q=0.1$  and $R_{sh}=5.14804M$).}
\begin{ruledtabular}
\begin{tabular}{cccc}
\vspace{-0.3cm}\\
$\ell$ &$R_{sh}^{(WKB-PT)}/M$ & $R_{sh}^{(Ei)}/M$ &$R_{sh}^{(Ch)}/M$
\vspace{0.1cm}\\ 
\hline
5& 4.63095 & 5.14804 & 4.68141\\ 
10& 4.88864 & 5.14804 & 4.90434\\ 
50& 5.09643 & 5.14804 & 5.09857\\ 
100& 5.12227 & 5.14804 & 5.12393\\ 
500& 5.14289 & 5.14804 & 5.14441\\ 
1000& 5.14547 & 5.14804 & 5.14698\\ 
5000& 5.14753 & 5.14804 & 5.14904\\ 
10000& 5.14779 & 5.14804 & 5.14929\\
50000& 5.14799 & 5.14804 & 5.14950\\
100000& 5.14802 & 5.14804 & 5.14953\\
500000& 5.14804 & 5.14804 & 5.14955\\
\end{tabular}
\end{ruledtabular}
\end{table}

Using equation \eqref{Shadow1}, a CEGB BH with $\alpha=0.1$ and $Q=0.1$ has a shadow with radius $R_{sh}=5.14804M$. In Table \ref{tab:table5} we show the results obtained for the shadow radius for the CEGB BH  for the fundamental mode ($n=0 $) and using various values of $\ell$. There, $R_{sh}^{(WKB-PT)}$ represents the shadow of the BH obtained from the WKB and PT potential methods, $R_{sh}^{(Ei)}$ denotes the BH shadow radius obtained by using the eikonal  equation and finally, the fourth column shows the results of Churilova's analytical equation. It should be noted that $R_{sh}^{(WKB-PT)}$ summarizes all the results of the WKB approximation methods at $1^{st}$, $3^{rd}$ and $6^{th }$ order and the PT potential because, as discussed in Table \ref{tab:table1}, these methods return the same values of the frequencies of the QNMs of the scalar and electromagnetic field in the eikonal limit.  Therefore, as an initial conclusion of this work we see that all the results obtained from the radius of its shadow show verify that the CEGB BH does indeed fulfill the connection between geometrical paramters and  the QNMs. The results given by Churilova's analytical eikonal approach show that $R_{sh}^{(Ch)}$ achieves values that are very close to those obtained by the other approaches. For example, for $\ell=500000$ the deviation error between $R_{sh}^{(WKB-PT)}$ and $R_{sh}^{(Ch)}$ is less than 1\%. 

For practical purposes, doing $\ell \rightarrow \infty$ removes  the dependency on $\ell$ from equation \eqref{Shadow4}, simplifying the CEGB BH shadow radius to
\begin{equation}
    R_{sh}^{(CEGB)}=\frac{486 \sqrt{3}  M^5}{ 162 M^4+3 M^2 \left(4 \alpha +9 Q^2\right)-4 \alpha  Q^2}
    \label{Shadow5}
\end{equation}

From this expression, the shadow radius for the limiting paraticular cases of the CEGB BH can be calculated. For example, taking $\alpha \rightarrow 0$ gives
\begin{equation}
    R_{sh}^{(RN)}=\frac{18 \sqrt{3} M^3}{6 M^2+ Q^2},
    \label{Shadow6}
\end{equation}
while taking $Q \rightarrow 0$ produces
\begin{equation}
    R_{sh}^{(EGB)}=\frac{81 \sqrt{3}  M^3}{27 M^2+ 2 \alpha}.
    \label{Shadow7}
\end{equation}

Similarlyly, by choosing $Q \rightarrow 0$ and $\alpha \rightarrow 0$ we recover the well-known limit of the Schwarzschild BH shadow radius, $R_{sh}^{(Sch)}=3 \sqrt{3} M$.

In Figure \ref{fig:fig5}, the solid line summarizes the BH shadow radius calculated using the QNMs given by the WKB-PT method and the eikonal limit formula while the dotted line shows  the shadow radius from Churilova's eikonal approach. It also ilustrates the effetcs  of the electric charge of the BH and the GB coupling constant on the shadow radius of a CEGB BH. As both parameters increase, the shadow radius decreases, which agrees with previous studies \cite{Guo2020, Panah2020, Chen2021, Liu2022}. This implies that, based on the range of possible values for the BH geometric parameters as stated in equation \eqref{alpharange}, the CEGB BH has a smaller shadow radius compared to the EGB 4D BH and the RN BH, and these BHs have a smaller shadow radius than that of a Schwarzschild BH

Figure (\ref{fig:fig5}) also show that the curves overlap for small values of $\alpha$ and $Q$ and then, in this region the CEGB BH shadow radius obtained from Churilova's eikonal approach is very useful due to its analytical simplicity, providing accurate results.

Fianlly, it is important to say that, in \cite{Guo2020}, the authors examine a hypothesis regarding a series of inequalities involving multiple parameters associated with the size of an EGB 4D BH. The proposal is that 
\begin{equation}
    \frac{3}{2} r_{+} \leq R_{ps} \leq \frac{1}{\sqrt{3}} R_{sh} \leq 3M,
\end{equation}
keeping in mind the ranges of equation \eqref{alpharange} for the BH geometric parameters. We have evaluated the validity of this hypothesis involving  the parameters that describe the size of a CEGB BH and our findings indicate that it satisfies the inequalities.

\section{Conclusion}

Our study revealed the relation between geometrical properties as the photon sphere or the shadow raidus and the eikonal QNMs of a CEGB BHs. The WKB method at different orders of approximation and the PT potential method have been shown to give the same results for the eikonal QNMs for both scalar and electromagnetic field perturbations. Our findings demonstrate a correlation between the real part of the eikonal QNM frequencies and the unstable circular null geodesic, as well as with the shadow radius of a CEGB BH. Furthermore, the real part of QNMs, the photon sphere, and shadow radius were found to be lower in CEGB BHs compared to those in EGB 4D, RN, and Schwarzschild BHs. A number of approximate equations for the radius of the photon sphere and shadows have been derived. In particular, Churilova's eikonal approach is valuable as it provides a simple, approximate and analytical equation for the CEGB BH shadow radius which  make other studies easier, such as the investigations on the observable properties related to astrophysical BH shadows. Hence, it is recommended to extend this approach to other BH solutions in order to test the methodology.  It would also be interesting to explore the potential of similar analytical approaches in spacetime arrangements that are not limited to being static, spherically symmetrical or asymptotically flat.

There are several topics for future research related to this work. The connections between the photon sphere or BH shadow and eikonal QNMs should be explored for BH solutions that are similar to the CEGB BH, such as those in alternative gravity theories . A particularly important BH solution to test these connections, from an astrophysical perspective, is the rotating CEGB BH, as recent theories also suggest connections between light rays, shadow radius, and eikonal QNMs in rotating spacetimes \cite{Jusufi2020-2, Yang2021}. The correspondence between eikonal QNMs and photon orbits of charged rotating BHs in GR has also been verified \cite{Cheng2021}. Also, in \cite{Kumar2020-5}, it was shown that the shadow of a rotating EGB 4D BH aligns with the observed characteristics of M87* BH shadow observed by the Event Horizon Telescope. It was found, for instance, that for a spin parameter of $a=0.1M$, the GB coupling constant must be $\alpha \leq 0.00394M^2$ (a very small value for $\alpha$ which matches a good approximation of Churilova's eikonal approach). Therefore, it would be beneficial to also provide an analytical approach to these problems, linking the eikonal QNMs to the geometric properties of rotating CEGB BHs.

\section*{Acknowledgements}
The authors acknowledge partial financial support from
Dirección de Investigación-Sede Bogotá, Universidad
Nacional de Colombia (DIB-UNAL) under HERMES Project
No. 57057 and Grupo de Astronomía, Astrofísica y
Cosmología-Observatorio Astronómico Nacional.

\end{document}